\documentclass[multphys,vecphys]{svmult}

\usepackage{graphicx}% Include figure files
\usepackage{dcolumn}% Align table columns on decimal point
\usepackage{bm}% bold math
\usepackage{makeidx}     % allows index generation
\usepackage{multicol}    % used for the two-column index
\usepackage{url}

\newcommand{\Li}{\mathrm{Li}}
\newcommand{\mean}[1]{\langle #1 \rangle}

\begin{document}

\title*{Information Dynamics in the Networked World}
\author{Bernardo A. Huberman \and
Lada A. Adamic}
\institute{HP Labs, 1501 Page Mill Road, CA 94304-1126
\url{huberman@hpl.hp.com}}
\maketitle

\begin{abstract}
We review three studies of information flow in social networks
that help reveal their underlying social structure, how
information spreads among them and why small world experiments
work.
\end{abstract}
\section{Introduction}

The problem of information flows in social organizations is
relevant to issues of productivity, innovation and the sorting out
of useful ideas from the general chatter of a community. How
information spreads determines the speed with which individuals
can act and plan their future activities. Moreover, information
flows take place within social networks whose nature is sometimes
difficult to establish. This is because the network itself is
sometimes different from what one would infer from the formal
structure of the group or organization.

The advent of email as the predominant means of communication in
the information society now offers a unique opportunity to observe
the flow of information along both formal and informal channels.
Not surprisingly, email has been established as an indicator of
collaboration and knowledge exchange
\cite{wellman02computersocial, whittaker96email, guimera02email,
tyler03email,eckmann03email}. Email is also a good medium for
research because it provides plentiful data on personal
communication in an electronic form. This volume of data enables
the discovery of shared interests and relationships where none
were previously known \cite{schwartz93discovering}.

In this chapter we will review three studies that utilized
networks exposed by email communication. In all three studies, the
networks analyzed were derived from email messages sent through
the Hewlett Packard Labs email server over the period of several
months in 2002 and 2003. The first study, by Tyler et al.
\cite{tyler03email}, develops an automated method applying a
betweenness centrality algorithm to rapidly identify communities,
both formal and informal, within the network. This approach also
enables the identification of leadership roles within the
communities. The automated analysis was complemented by a
qualitative evaluation of the results in the field.

The second study, by Wu et al. \cite{wu03flow} analyzes email
patterns to model information flow in social groups, taking into
account the observation that an item relevant to one person is
more likely to be of interest to individuals in the same social
circle than those outside of it. This is due to the fact that the
similarity of node attributes in social networks decreases as a
function of the graph distance. An epidemic model on a scale-free
network with this property has a finite threshold, implying that
the spread of information is limited. These predictions were
tested by measuring the spread of messages in an organization and
also by numerical experiments that take into consideration the
organizational distance among individuals.

Since social structure affects the flow of information, knowledge
of the communities that exist within a network can also be used
for navigating the networks when searching for individuals or
resources. The study by Adamic and Adar\cite{adamic03search}, does
just this, by simulating Milgram's small world experiment on the
HP Labs email network. The small world experiment has been carried
out a number of times over the past several decades, each time
demonstrating that individuals passing messages to their friends
and acquaintances can form a short chain between two people
separated by geography, profession, and race. While the existence
of these chains has been established, how people are able to
navigate without knowing the complete social networks has remained
an open question. Recently, models have been proposed to explain
the phenomenon, and the work of Adamic and Adar is a first study
to test the validity of these models on a social network.

\section{Email as Spectroscopy \label{spectroscopy}}
Communities of practice are the informal networks of collaboration
that naturally grow and coalesce within and outside organizations.
Any institution that provides opportunities for communication
among its members is eventually threaded by communities of people
who have similar goals and a shared understanding of their
activities \cite{ouchi80markets}. These communities have been the
subject of much research as a way to uncover the reality of how
people find information and execute their tasks. (for example, see
\cite{blau63organizations,burt80network,wasserman94socialnetwork},
or for a survey see \cite{scott92organizations}).

These informal networks coexist with the formal structure of the
organization and serve many purposes, such as resolving the
conflicting goals of the institution to which they belong, solving
problems in more efficient ways \cite{huberman95communities}, and
furthering the interests of their members. Despite their lack of
official recognition, informal networks can provide effective ways
of learning, and with the proper incentives actually enhance the
productivity of the formal organization
\cite{crozier64bureaucratic,crane72colleges, lave91learning}.

Recently, there has been an increased amount of work on
identifying communities from online interactions (a brief overview
of this work can be found in \cite{wellman02computersocial}). Some
of this work finds that online relationships do indeed reflect
actual social relationships, thus adding effectively to the
``social capital'' of a community. Ducheneaut and Bellotti
\cite{ducheneaut02email} conducted in-depth field studies of email
behavior, and found that membership in email communities is quite
fluid and depends on organizational context. Mailing lists and
personal web pages also serve as proxies for social relationships
\cite{adamic03friends}, and the communities identified from these
online proxies resemble the actual social communities of the
represented individuals. Because of the demonstrated value of
communities of practice, a fast, accurate method of identifying
them is desirable.

Classical practice is to gather data from interviews, surveys, or
other fieldwork and to construct links and communities by manual
inspection (see \cite{allen84flow,hinds95communication} or an
Internet-centric approach in \cite{garton97online}).  These
methods are accurate but time-consuming and labor-intensive,
prohibitively so in the context of a very large organization.
Alani et al. \cite{alani02practice} recently introduced a
semi-automated utility that uses a simple algorithm to identify
nearest neighbors to one individual within a university
department.

The method of Tyler et al. \cite{tyler03email} uses email data to
construct a network of correspondences, and then discovers the
communities by partitioning this network. It was applied to a set
of over one million email messages collected over a period of
roughly two months at HP Labs in Palo Alto, an organization of
approximately 400 people. The only pieces of information used from
each email are the names of the sender and receiver (i.e., the
``to:'' and ``from:'' fields), enabling the processing of a large
number of emails while minimizing privacy concerns.

The method was able to identify small communities within the
organization, and the leaders for those communities, in a matter
of hours, running on a standard Linux desktop PC. This experiment
was followed by a qualitative evaluation of the experimental
results in the ``field'', which consisted of sixteen face-to-face
interviews with individuals in HP Labs.  The interviews validated
the results obtained by the automated process, and provided
interesting perspectives on the communities identified. We
describe the results in more detail below.

\subsection{Identifying Communities}
It is straightforward to construct a graph based on email data, in
which vertices represent people and edges are added between people
who exchanged at least a threshold number of email messages. Next,
one can identify communities: subsets of related vertices, with
many edges connecting vertices of the same subset, but few edges
lying between subsets \cite{girvan02community}.

The method of Wilkinson and Huberman \cite{wilkinson02genes},
related to the algorithm of Girvan and Newman
\cite{girvan02community}, partitions a graph into discrete
communities of nodes and is based on the idea of betweenness
centrality, or betweenness, first proposed by Freeman
\cite{freeman77centrality}. The betweenness of an edge is defined
as the number of shortest paths that traverse it. This property
distinguishes inter-community edges, which link many vertices in
different communities and have high betweenness, from
intra-community edges, whose betweenness is low.

\begin{figure}[tb]
  \centering\includegraphics[scale=1]{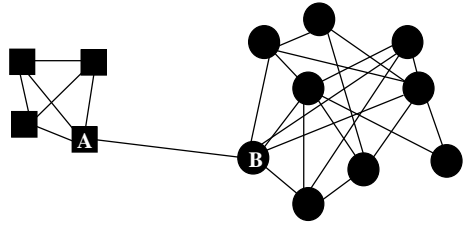}
  \caption{An example graph with edge AB having high betweenness.\label{wheretocut}}
\end{figure}

To illustrate the community discovery process, consider the
 small graph shown in Figure \ref{wheretocut}. This
graph consists of two well-defined communities: the four vertices
denoted by squares, including vertex A, and the nine denoted by
circles, including vertex B. Edge AB has the highest betweenness,
because all paths between any circle and square must pass through
it. If one were to remove it, the squares and circles would be
split into two separate communities. The algorithm of Wilkinson et
al. repeatedly identifies inter-community edges of large
betweenness such as AB and removes them, until the graph is
resolved into many separate communities.

\begin{figure}
 \centering\includegraphics[scale=1]{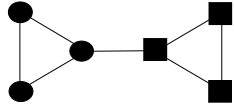}
 \caption{The smallest possible graph of two viable communities.\label{twocommunities}}
\end{figure}

Because the removal of an edge strongly affects the betweenness of
many others, the values were repeatedly updated with the fast
algorithm of Brandes
\cite{brandes01betweenness,newman01scientist,girvan02community}.
The procedure stops removing edges when it cannot further
meaningfully subdivide communities. Figure \ref{twocommunities}
shows the smallest possible component that can be subdivided into
two viable subcommunities. It has 6 nodes, consisting of two
triangles linked by one edge. A component with fewer than 6 nodes
cannot be subdivided further.

\begin{figure}
  \centering\includegraphics[scale=1]{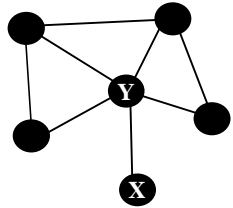}
  \caption{An example graph of one community that does not contain distinct sub-communities.\label{leafbetween}}
\end{figure}

Components of size $\geq 6$, for example the group of size nine in
Figure \ref{wheretocut}, can also constitute single cohesive
communities. Figure \ref{leafbetween} shows how the algorithm
determines when to stop subdividing a community. The edge XY has
the highest betweenness, but removing it would separate a single
node, which does not constitute a viable community. In general,
the single edge connecting a leaf vertex (such as X in Figure
\ref{leafbetween}) to the rest of a graph of $N$ vertices has a
betweenness of $N-1$ , because it contains the shortest path from
X to all $N-1$  other vertices. The stopping criterion for
components of size $\geq$ 6 is therefore that the highest
betweenness of any edge in the component be equal to or less than
$N-1$.

\subsection{Multiple Community Structures}

As mentioned above, the removal of any one edge affects the
betweenness of all the other edges, particularly in large,
real-world graphs such as the email graph. Early in the process,
there are many inter-community edges which have high betweenness
and the choice of which to remove, while arbitrary, dictates which
edges will be removed later. For example, a node belonging to two
communities can be placed in one or the other by the algorithm,
depending on the order in which edges are removed. One can take
advantage of this arbitrariness to repeatedly partition the graph
into many different ``structures'' or sets of communities. These
sets are then compared and aggregated into a final list of
communities.

Wilkinson and Huberman \cite{wilkinson02genes} introduced
randomness into the algorithm by calculating the shortest paths
from a random subset as opposed to all the nodes. The algorithm
cycles randomly through at least $m$ centers (where $m$ is some
cutoff) until the betweenness of at least one edge exceeds the
threshold betweenness of a ``leaf'' vertex. The edge whose
betweenness is highest at that point is removed, and the procedure
is repeated until the graph has been separated into communities.
The modified algorithm may occasionally remove an intra-community
edge, but such errors are unimportant when a large number of
structures is aggregated.

Applying this modified process $n$ times yields $n$ community
structures imposed on the graph. One can then compare the
different structures and identify communities. For example, after
imposing 50 structures on a graph, one might find: a community of
people A, B, C, and D in 25 of the 50 structures; a community of
people A, B, C, D, and E in another 20; and one of people A, B, C,
D, E and F in the remaining 5. This result is reported in the
following way: A(50) B(50) C(50) D(50) E(25) F(5) which signifies
that A, B, C, and D form a well-defined community, E is related to
this community, but also to some other(s), and F is only slightly,
possibly erroneously, related to it. For details of the
aggregation procedure, please see \cite{wilkinson02genes}.

The entire process of determining community structure within the
graph is displayed below.

\begin{itemize}
\item For $i$ iterations, repeat \{
\begin{enumerate}
\item Break the graph into connected components.
\item For each component, check to see if component is a community.
\begin{itemize}
\item If so, remove it from the graph and output it.
\item If not, remove edges of highest betweenness, using the
modified Brandes algorithm for large components, and the normal
algorithm for small ones. Continue removing edges until the
community splits in two.
\end{itemize}
\item Repeat step 2 until all vertices have been removed from the
graph in communities. \} \end{enumerate}

\item Aggregate the $i$ structures into a final list of communities.
\end{itemize}

\subsection{Results}
The algorithm was applied to email data from the HP Labs mail
server from the period November 25, 2002 to February 18, 2003,
with 185,773 emails exchanged between the 485 HP Labs employees.
For simplicity, emails that had an external origin or destination
were omitted. Messages sent to a list of more than 10 recipients
were likewise removed, as these emails were often lab-wide
announcements (rather than personal communication), which were not
useful in identifying communities of practice.

A graph was constructed from this data by placing edges between
any two individuals that had exchanged at least 30 emails in
total, and at least 5 in both directions. The threshold eliminated
infrequent or one-way communication, and eliminated some
individuals from the graph who either sent very few emails or used
other email systems.  The resulting graph consisted of 367 nodes,
connected by 1110 edges.

There was one giant connected component of 343 nodes and six
smaller components ranging in size from 2 to 8. The modified
Brandes algorithm detected 60 additional distinct communities
within the giant component. The largest community consisted of 57
individuals, and there were several communities of size 2. The
mean community size was 8.4, with standard deviation 5.3. A
comparison of these communities with information from the HP
corporate directory revealed that 49 of the 66 communities
consisted of individuals entirely within one lab or organizational
unit. The remaining 17 contained individuals from two or more
organizations within the company.

\subsection{Identifying Leadership Roles}
\begin{figure}
  \centering\includegraphics[scale=0.75]{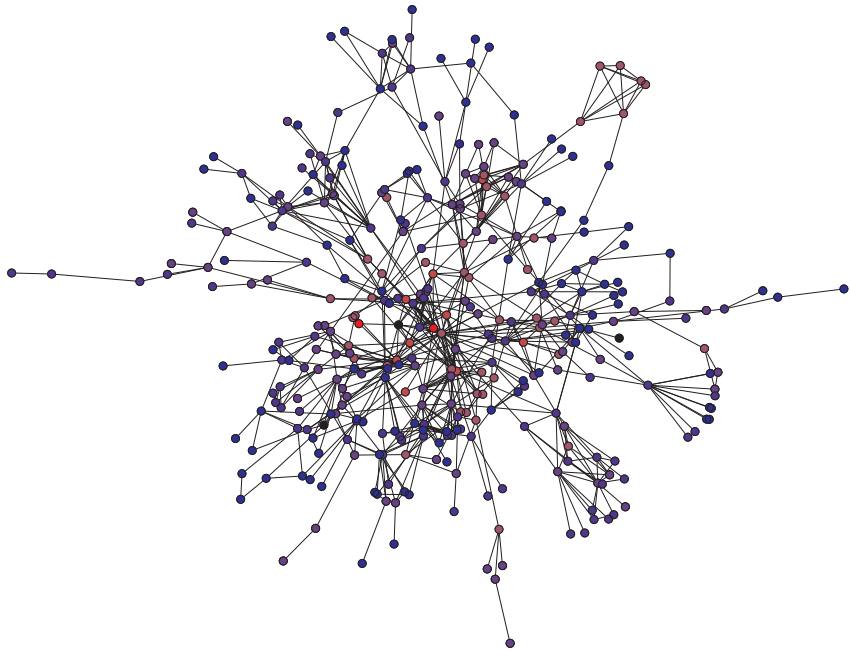}
  \caption{The giant connected component of the HP Labs email network.
  The redness of a vertex indicates an individual's closeness to the top of the lab hierarchy
  (red-close to top, blue-far from top, black-no data available).\label{springlayout}}
\end{figure}

In addition to identifying formal and informal work communities,
it is also possible to draw inferences about the leadership of an
organization from its communication data. One method is to
visualize the above graph of the HP Labs email network with a
standard force-directed spring algorithm
\cite{fruchterman91graph}, shown in Figure \ref{springlayout}.
This spring layout of the email network does not use any
information about the actual organization structure, and yet high
level managers (the reddest nodes are at the top of the hierarchy)
are placed close to the center of the graph. The trend is
quantified in Table \ref{distvsdepth}, which lists the average
hierarchy depth (levels from the lab director) as a function of
the position in the layout from the center.

Note that there is a group of 6 nodes in the upper right portion
of the graph that are quite removed from the center, but are
relatively high in the organizational hierarchy. This is the
university relations group that reports directly to the head of HP
Labs, but has no other groups reporting to it. Hence the layout
algorithm correctly places them on the periphery of the graph,
since their function, that of managing HP's relationship with
universities, while important, is not at the core of day-to-day
activities of the labs.

\begin{table}[tbp]
\begin{center}
\begin{tabular}{|c|c|c|} \hline
distance from center & number of vertices & average depth in
hierarchy \\ \hline
$< 0.1$    & 14   &   2.6 \\
0.1 to 0.2 &   32   &   3.0 \\
0.2 to 0.3 & 56 & 3.2 \\
0.3 to 0.4 &  66 & 4.0 \\
0.4 to 0.5 & 56 & 4.0 \\
0.5 to 0.6 &  45 & 4.2 \\
0.6 to 0.7 & 42 & 4.0 \\
0.7 to 0.8 & 12 & 3.9 \\
0.8 to 0.9 & 13 & 3.8 \\ \hline
\end{tabular}
\end{center}
\caption{Average hierarchy depth by distance from center in
layout} \label{distvsdepth}
\end{table}

Evaluating communication networks with this technique could
provide information about leadership in communities about which
little is known.  Sparrow proposed this approach for analyzing
criminal networks \cite{sparrow91criminal}, noting that
``Euclidean Centrality is probably the closest to the reality'' of
the current criminal network analysis techniques. More recently,
Krebs applied centrality measures and graphing techniques
\cite{krebs02terrorist} to the terrorist networks uncovered in the
9/11 aftermath. He found that the average shortest path was
unusually long for such a small network, and concluded that the
operation had traded efficiency for secrecy - individuals in one
part of the network did not know those in other parts of the
network. If one cell had been compromised, the rest of the network
would remain relatively unaffected. Several social network
centrality measures pointed to Mohamed Atta's leadership role in
the attacks of Sept. 11. The role was also confirmed by Osama bin
Laden in a video tape following the attacks.

\subsection{Field Evaluation}
The HP Labs social network, being much less covert, could readily
be compared to the structure of the formal organization.
Nevertheless, the informal communities identified by the algorithm
could not be verified in this way. Tyler et al. decided to
validate the results of their algorithm by conducting a brief,
informal field study. Sixteen individuals chosen from seven of the
sixty communities identified were interviewed informally. The
communities chosen represented various community sizes and levels
of departmental homogeneity. They ranged in size from four to
twelve people, and three out of the seven were heterogeneous
(included members of at least two different departmental units
within the company).

All sixteen subjects gave positive affirmation that the community
reflected reality.  More specifically, eleven described the group
as reflecting their department, four described it as a specific
project group, and one said it was a discussion group on a
particular topic. Nine of the sixteen (56.25\%) said nobody was
missing from the group, six people (37.5\%) said one person was
missing, and one person (6.25\%) said two people were missing.
Conversely, ten of the sixteen (62.5\%) said that everybody in the
group deserved to be there, whereas the remaining six (37.5\%)
said that one person in the group was misclassified.

The interviews confirmed that most of the communities identified
were based on organization structure.  However, the communities
also tended to include people who were de facto department
members, but who did not technically appear in the department's
organization chart, such as interns or people whose directory
information had changed during the two months of the study.
Finally, the algorithm seemed to succeed in dividing departmental
groups whose work is distinct, but lumped together groups whose
projects overlap.

Heterogeneous, cross-department communities are of particular
interest because they cannot be deduced from the formal
organization. The interviews revealed that most of them
represented groups formed around specific projects, and in one
case, a discussion forum. For example, one community contained
three people from different labs coordinating on one project: a
technology transfer project manager, a researcher who was the
original designer of a piece of PC hardware, and an engineer
redesigning the hardware for a specific printer.

\subsection{Discussion}
The power of this method for identifying communities and
leadership is in its automation.  It does an effective job of
uncovering communities of practice with nothing more than email
log (``to:'' and ``from:'') data.  Its simplicity means that it
can be applied to organizations of thousands and produce results
efficiently. However, it is important for computing centrality
measures to be able to define membership in an organization as
well as disambiguate identities. In a setting like a corporate
lab, the organization is clearly defined and identities can be
clarified from official directories. In an informal network,
however, these tasks are much more difficult.

Communities identified in this automated way lack the richness in
contextual description provided by ethnographic approaches. They
do not reveal the nature or character of the identified
communities, the relative importance of one community to another,
or the subtle inter-personal dynamics within the communities.
These kinds of details can only be uncovered with much more data-
or labor-intensive techniques. However, in cases where an
organization is very large, widely dispersed, or incompletely
defined (informal), this method provides an suitable alternative
or compliment to the more traditional, labor-intensive approaches.

\section{Information Flow in Social Groups \label{infoflow}}
In the previous section we saw that individuals tend to organize
both formally and informally into groups based on their common
activities and interests. In this section we examine how this
structure in the interaction network affects the way information
spreads. This is not unlike the transmission of an infectious
agent among individuals, where the pattern of contacts determines
how far a disease spreads. Thus one would expect that epidemic
models on graphs are relevant to the study of information flow in
organizations. In particular, recent work on epidemic propagation
on scale free networks found that the threshold for an epidemic is
zero, implying that a finite fraction of the graph becomes
infected for arbitrarily low transmission probabilities
\cite{zoltan02halting,pastor-satorras01epidemic,newman02emailnetworks}.
The presence of additional network structure was found to further
influence the spread of disease on scale-free graphs
\cite{eguiluz02epidemicclust,vazquezPRE2003,newman02assortative}.

\begin{figure}[tbp]
\begin{center}
\includegraphics[scale=0.4]{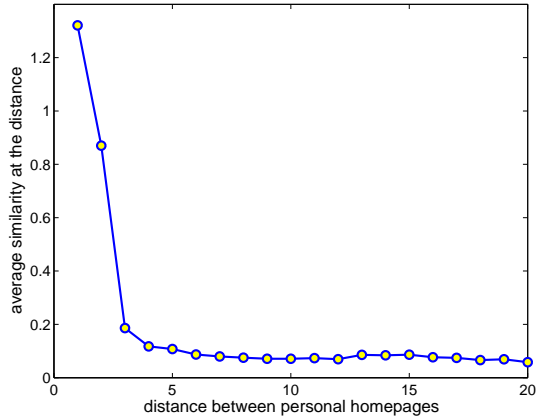}
\end{center}
\caption[Node similarity as a function of distance in the graph
]{Average similarity of Stanford student homepages as a function
of the number of hyperlinks separating them.} \label{disttolikeav}
\end{figure}

There are, however, differences between information flows and the
spread of viruses. While viruses tend to be indiscriminate,
infecting any susceptible individual, information is selective and
passed by its host only to individuals the host thinks would be
interested in it. The information any individual is interested in
depends strongly on their characteristics. Furthermore,
individuals with similar characteristics tend to associate with
one another, a phenomenon known as homophily
\cite{lazarsfeld54friendship, touhey74similarity,feld81social}.
Conversely, individuals many steps removed in a social network on
average tend not to have as much in common, as shown in a study
\cite{adamic03friends} of a network of Stanford student homepages
and illustrated in Figure \ref{disttolikeav}.

Wu et al. \cite{wu03flow} introduced an epidemic model with decay
in the transmission probability of a particular piece of
information as a function of the distance between the originating
source and the current potential target. This epidemic model on a
scale-free network has a finite threshold, implying that the
spread of information is limited. The predictions were further
tested by observing the prevalence of messages in an organization
and also by numerical experiments that take into consideration the
organizational distance among individuals.

Consider the problem of information transmission in a power-law
network whose degree distribution is given by \begin{equation} p_k
= Ck^{-\alpha} e^{-k/\kappa},\end{equation} where $\alpha>1$,
there is an exponential cutoff at $\kappa$ and $C$ is determined
by the normalization condition. A real world graph will at the
very least have cutoff at the maximum degree $k=N$, where $N$ is
the number of nodes, and many networks show a cutoff at values
much smaller than $N$. For the analysis of the spread of
information flow on networks, Wu et al. used generating functions,
whose application to graphs with arbitrary degree distributions is
discussed in \cite{newman01graphs}.  For a power-law network the
generating function is given by
\begin{equation} G_0(x) = \sum_{k=1}^\infty p_k x^k =
\frac{\Li_\alpha(x e^{-k/\kappa})}{\Li_\alpha(e^{-1/\kappa})}.
\end{equation}
where $Li_n(x)$ is the $n$th polylogarithm of $x$.

Following the analysis in \cite{newman02epidemic} for the SIR
(susceptible, infected, removed) model, one can estimate the
probability $p_m^{(1)}$ that the first person in the community who
has received a piece of information will transmit it to $m$ of
their neighbors. Using the binomial distribution, we find
\begin{equation} p_m^{(1)} = \sum_{k=m}^\infty p_k {k \choose m}
T^m (1-T)^{k-m},
\end{equation} where the superscript
``$(1)$'' refers to first neighbors, those who received the
information directly from the initial source. The
\emph{transmissiblity} $T$ is the average total probability that
an infective individual will transmit an item to a susceptible
neighbor and is derived in \cite{newman02epidemic} as a function
of $r_{ij}$, the rate of contacts between two nodes, and $\tau_i$,
the time a node remains infective. If $r_{ij}$ and $\tau_i$ are
iid randomly distributed according to the distributions $P(r)$ and
$P(\tau)$, then the item will propagate as if all transmission
probabilities are equal to a constant $T$.

\begin{equation}
T= \langle T_{ij} \rangle = 1 - \int_{0}^\infty  dr d\tau
P(r)P(\tau)e^{-r \tau} \end{equation}

The generating function for $p_m^{(1)}$ is given by
\begin{eqnarray}
G^{(1)} (x) &=& \sum_{m=0}^{\infty} \sum_{k=m}^\infty p_k {k
\choose m} T^m (1-T)^{k-m} x^m\\
&=& G_0(1+(x-1)T) = G_0(x;T).
\end{eqnarray}

Suppose the transmissibility decays as a power of the distance
from the initial source. The probability that an $m$th neighbor
will transmit the information to a person with whom he has contact
is given by
\begin{equation} T^{(m)} =
(m+1)^{-\beta}T,
\end{equation}
where $\beta>0$ is the decay constant. $T^{(m)}=T$ at the
originating node ($m=0$) and decays to zero as
$m\rightarrow\infty$. Power-law decay is the weakest form of decay
and the results obtained from it will also be valid for stronger
functional forms such as an exponential decay.

The generating function for the transmission probability to 2nd
neighbors can be written as
\begin{equation} G^{(2)}(x) = \sum_k
p_k^{(1)} [G_1^{(1)}(x)]^k = G^{(1)} (G_1^{(1)}(x)),
\end{equation} where \begin{equation} G_1^{(1)}(x) = G_1(x;
2^{-\beta}T) = G_1(1+ (x-1) 2^{-\beta}T) \end{equation} and
\begin{equation} G_1(x)=\frac{\sum_k k p_k x^k}{x \sum_k k
p_k}=\frac{G_0'(x)}{G_0'(1)} \end{equation} is the generating
function of the degree distribution of a vertex reached by
following a randomly chosen edge, not counting the edge itself
\cite{newman01graphs}. Similarly, if we define $G^{(m)}(x)$ to be
the the generating function for the number of $m$th neighbors
affected, then we have
\begin{equation} G^{(m+1)}(x) = G^{(m)} (G_1^{(m)}(x)) \quad
\mbox{for }m\ge 1, \end{equation} where
\begin{equation} G_1^{(m)}(x) = G_1(x; (m+1)^{-\beta}T) = G_1(1+
(x-1)(m+1)^{-\beta}T). \end{equation}  Or, more explicitly,
\begin{equation} G^{(m+1)}(x) = G^{(1)} ( G_1^{(1)} ( G_1^{(2)} (
\cdots G_1^{(m)} (x)))). \end{equation}

The average number $z_{m+1}$ of $(m+1)$th neighbors is
\begin{equation} z_{m+1} = {G^{(m+1)}}'(1) = {G_1^{(m)}}'(1)
{G^{(m)}}'(1) = {G_1^{(m)}}'(1) z_m. \end{equation}

The condition that the size of the outbreak remains finite is that
at some distance $m+1$, fewer individuals will be infected than at
distance $m$, i.e. the spread of the infection is halting. This
can be expressed as
\begin{equation} \frac{z_{m+1}} {z_m} = {G_1^{(m)}}'(1) < 1, \end{equation} or
\begin{equation} (m+1)^{-\beta}T G_1'(1) < 1.
\end{equation}
Note that $G_1'(1)$ does not diverge when $\alpha < 3$ due to the
presence of a cutoff at $\kappa$. For any decaying $T$, the left
hand side of the inequality above goes to zero when $m\to \infty$,
so the condition is eventually satisfied for large $m$. Therefore
the average total size
\begin{equation} \mean s = \sum_{m=1}^\infty z_m \end{equation} is always
finite if the transmissibility decays with distance.

\begin{figure}
  \centering\includegraphics[scale=0.45]{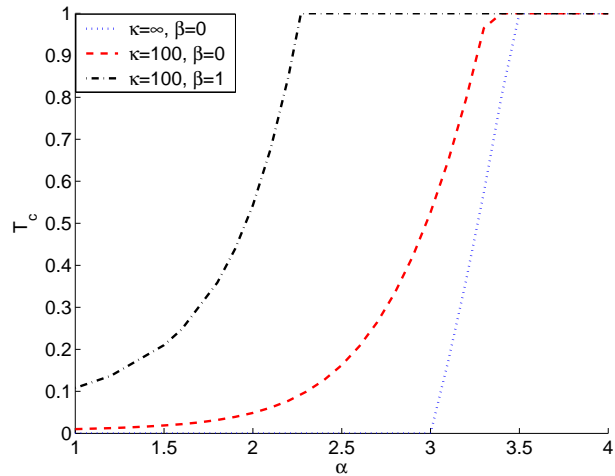}
  \caption{\label{threshold} $T_c$ as a function of $\alpha$. The three different curves,
  from bottom to top are: 1) no decay in transmission probability, no exponential cutoff in the
  degree distribution ($\kappa=\infty, \beta=0$). 2) $\kappa=100,
  \beta=0$, 3) $\kappa=100, \beta=1$.}
\end{figure}

Wu et al. compared their model with previous
results~\cite{pastor-satorras01epidemic} on disease spread on
scale-free networks, by considering a network made up of $10^6$
vertices. An epidemic was defined to be an outbreak affecting more
than 1\% or $10^4$ vertices. Thus for fixed $\alpha, \kappa$ and
$\beta$, $T_c$ is the critical transmissibility above which $\mean
s$ would be made to exceed $10^4$.

The numerical result of $T_c$ as a function of $\alpha$ is shown
in Figure $\ref{threshold}$.  When $\beta=0$ (there is no decay in
transmission probability), $\kappa=\infty$ (there is no cutoff in
the degree distribution), and $\alpha < 3$, $T_c$ is zero and
epidemics encompassing more than $10^4$ vertices occur for
arbitrarily small $T$, as was found in
\cite{pastor-satorras01epidemic}. Keeping $\beta$ at zero and
adding a cutoff at $\kappa=100$ produces a non-zero critical
transmissibility $T_c$, as was found in \cite{newman02epidemic}.
For $\alpha=2$, a typical value for real-world networks, $T_c$ is
still very near zero, meaning that for most values of $T$,
epidemics do occur. However, when we impose a decay in
transmissibility by setting $\beta$ to 1, $T_c$ rises
substantially. For example, $T_c$ jumps to 0.54 at $\alpha=2$ and
rises rapidly to 1 as $\alpha$ increases further, implying that
the information may not spread over the network.

\begin{figure}[tbp]
\begin{center}
\includegraphics[scale=0.45]{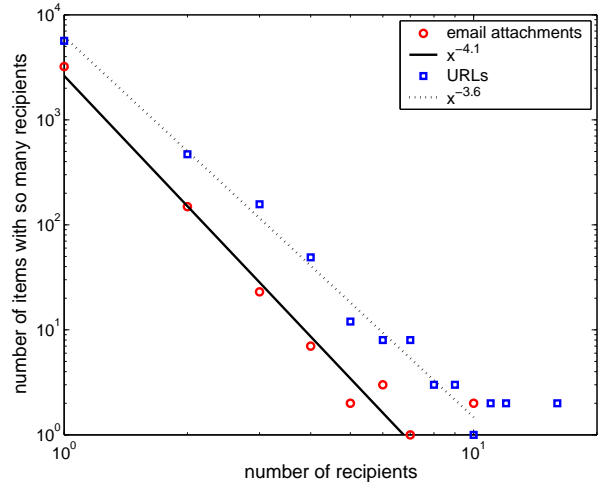}
\end{center}
\caption[Number of people receiving URLs and attachments]{Number
of people receiving URLs and attachments } \label{urlsattachments}
\end{figure}

In order to validate empirically that the spread of information
within a network of people is limited, and hence distinct from the
spread of a virus, a sample from the mail clients of 40
individuals (30 within HP Labs, and 10 from other areas of HP,
other research labs, and universities) was gathered.  Each
volunteer executed a program that identified URLs and attachments
in the messages in their mailboxes, as well as the time the
messages were received. This data was cryptographically hashed to
protect the privacy of the users.  By analyzing the message
content and headers, the data was restricted to include only
messages which had been forwarded at least one time, thereby
eliminating most postings to mailing lists and more closely
approximating true inter-personal information spreading behavior.
The median number of messages in a mailbox in the sample was 2200,
indicating that many users keep a substantial portion of their
email correspondence. Although some messages may have been lost
when users deleted them, it was assumed that a majority of
messages containing useful information had been retained.

Figure \ref{urlsattachments} shows a histogram of how many users
had received each of the 3401 attachments and 6370 URLs. The
distribution shows that only a small fraction (5\% of attachments
and 10\% of URLs) reached more than 1 recipient. Very few (41 URLs
and 6 attachments) reached more than 5 individuals, a number
which, in a sample of 40, starts to resemble an outbreak. In
follow-up discussions with the study subjects, the content and
significance of most of these messages was identified. 14 of the
URLs were advertisements attached to the bottom of an email by
free email services such as Yahoo and MSN. These are in a sense
viral, because the sender is sending them involuntarily. It is
this viral strategy that was responsible for the rapid buildup of
the Hotmail free email service user base.  10 URLs pointed to
internal HP project or personal pages, 3 URLs were for external
commercial or personal sites, and the remaining 14 could not be
identified.

\begin{figure}[tbp]
\begin{center}
\includegraphics[scale=0.45]{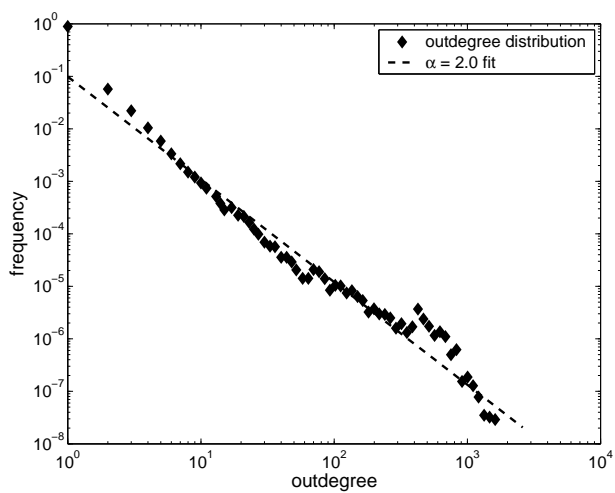}
\end{center}
\caption[Outdegree distribution for HP Labs email graph]{Outdegree
distribution for all senders (224,514 in total) sending email to
or from the HP Labs email server over the course of 3 months. The
outdegree of a node is the number of correspondents the node sent
email to.} \label{outdegdist}
\end{figure}

The next portion of the analysis analyzed the effect of decay in
the transmission probability on the email graph at HP Labs. The
graph was constructed from recorded logs of all incoming and
outgoing messages over a period of 3 months.  The graph has a
nearly power-law out degree distribution, shown in Figure
\ref{outdegdist}, including both internal and external nodes.
Because all of the outgoing and incoming contacts were recorded
for internal nodes, their in and out degrees were higher than for
the external nodes for which we could only record the email they
sent to and received from HP Labs. A graph with the internal and
external nodes mixed (as in \cite{ebel02email}) was used to
specifically demonstrate the effect of a decay on the spread of
email in a power-law graph.

The spread of a piece of information was simulated by selecting a
random initial sender to infect and following the email log
containing 120,000 entries involving over 7,000 recipients in the
course of a week. Every time an infective individual (one willing
to transmit a particular piece of information) was recorded as
sending an email to someone else, they had a constant probability
$p$ of infecting the recipient. Hence individuals who email more
often have a higher probability of infecting. It is also assumed
that an individual remains infective  for a period of 24 hours.

Next a decay was introduced in the one-time transmission
probability $p_{ij}$ as $p*d_{ij}^{-1.75}$, where $d_{ij}$ is the
distance in the organizational hierarchy between individuals $i$
and $j$. The exponent roughly corresponds to the decay in
similarity between homepages shown in Figure \ref{disttolikeav}.
Here $r_{ij} = p_{ij}*f_{ij}$, where $f_{ij}$ is the frequency of
communication between the two individuals, obtained from the email
logs. The decay represents the fact that individuals closer
together in the organizational hierarchy share more common
interests. Individuals have a distance of one to their immediate
superiors and subordinates and to those they share a superior
with. The distance between someone within HP labs and someone
outside of HP labs was set to the maximum hierarchical distance of
8.

\begin{figure}[tbp]
\begin{center}
\includegraphics[scale=0.4]{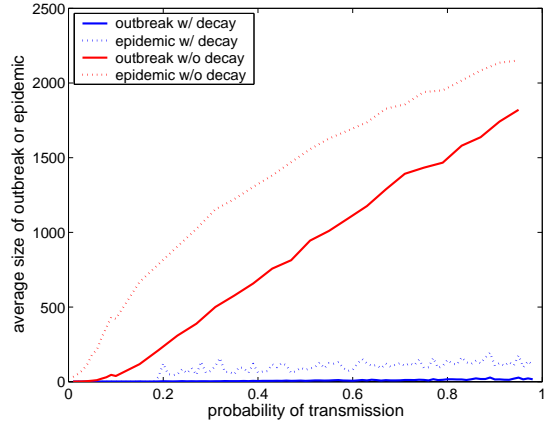}
\end{center}
\caption[Outbreak size as a function of the transmission
probability $p$ ]{Average outbreak and epidemic size as a function
of the transmission probability $p$. The total number of potential
recipients is 7119.} \label{outbreak}
\end{figure}

Figure \ref{outbreak} shows the variation in the average outbreak
size, and the average epidemic size (chosen to be any outbreak
affecting more than 30 individuals). Without decay, the epidemic
threshold falls below $p=0.01$. With decay, the threshold is set
back to $p = 0.20$ and the outbreak epidemic size is limited to
about 50 individuals, even for $p = 1$.

As these results show, the decay of similarity among members of a
social group has strong implications for the propagation of
information among them. In particular, the number of individuals
that a given email message reaches is very small, in contrast to
what one would expect on the basis of a virus epidemic model on a
scale free graph. The implication of this finding is that merely
discovering hubs in a community network is not enough to ensure
that information originating at a particular node will reach a
large fraction of the community.

\section{Small World Search}
In the preceding section we discussed how the tendency of like
individuals to associate with one another can affect the flow of
information within an organization. In this section we will show
how one can take advantage of the very same network structure to
navigate social ties and locate individuals.

The observation that any two people in the world are most likely
linked by a short chain of acquaintances, known as the ``small
world'' phenomenon has been the focus of much research over the
last forty years
\cite{milgram67,travers69smallworld,lundberg75smallworld,killworth78reverse}.
In the 1960's and 70's, articipants in small world experiments
successfully found paths from Nebraska to Boston and from Los
Angeles to New York. In an experiment in 2001 and 2002, 60,000
individuals were able to repeat the experiment using email to form
chains with just four links on average across different contents
\cite{dodds03networks}. The small world phenomenon is currently
exploited by commercial networking services such as LinkedIn,
Friendster, and
Spoke\footnote{\begin{flushleft}\url{http://www.linkedin.com/},
\mbox{\url{http://www.friendster.com}},
\mbox{\url{http://www.spokesoftware.com}}\end{flushleft}}to help
people network, for both business and social purposes.

The existence of short paths is not particularly surprising in and
of itself. Although many social ties are ``local'' meaning that
they are formed through ones work or place of residence, Watts and
Strogatz\cite{watts98smallworld} showed that it takes only a few
``random'' links between people of different professions or
location to create short paths in a social network and make the
world ``small''. In addition, Pool and Kochen\cite{pool78contacts}
have estimated that an average person has between 500 and 1,500
acquaintances.  Ignoring for the moment overlap in one's circle of
friends, one would have  $ ~ 1,000^2$ or $1,000,000$ friends of
friends, and $1,000^3$ or one billion
friends-of-friends-of-friends. This means that it would take only
2 intermediaries to reach a number of people on the order of the
population of the entire United States.

Although the existence of short paths is not surprising, it is
another question altogether how people are able to select among
hundreds of acquaintances the correct person to form the next link
in the chain. Killworth and Barnard\cite{killworth78reverse}
performed the ``reverse'' experiment to measure how many
acquaintances a typical person would use as a first step in a
small world experiment. Presented with 1,267 random targets, the
subjects chose about 210 different acquaintances on average, based
overwhelmingly on geographic proximity and similarity of
profession to the targets.

Recently, mathematical models have been proposed to explain why
people are able to find short paths. The model of Watts, Dodds,
and Newman \cite{watts2002search} assumes that individuals belong
to groups that are embedded hierarchically into larger groups. For
example an individual might belong to a research lab, that is part
of an academic department at a university, that is in a school
consisting of several departments, that is part of a university,
that is one of the academic institutions in the same country, etc.
The probability that two individuals have a social tie to one
another is proportional to $\exp^{-\alpha h}$, where $h$ is the
height of their lowest common branching point in the hierarchy.

\begin{figure}[tbp]
\begin{center}
\includegraphics[scale=0.4]{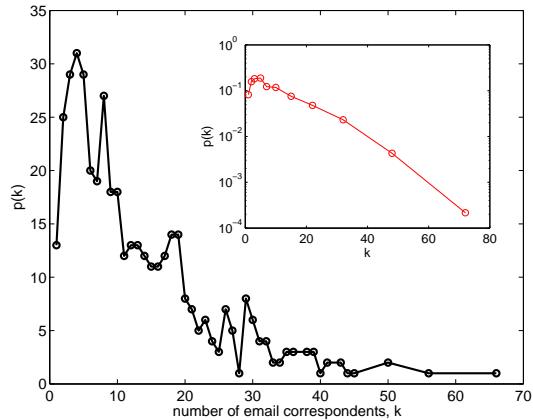}
\end{center}
\caption[Degree distribution in the HP Labs email network ]{Degree
distribution in the HP Labs email network. Two individuals are
linked if they exchanged at least 6 emails in either direction.
The inset shows the same distribution, but on a semilog scale, to
illustrate the exponential tail of the distribution }
\label{swlinkdist}
\end{figure}

The decay in linking probability means that two people in the same
research laboratory are more likely to know one another than two
people who are in different departments at a university. The model
assumes a number of separate hierarchies corresponding to
characteristics such as geographic location or profession. In
reality, the hierarchies may be intertwined, for example
professors at a university living within a short distance of the
university campus, but for simplicity, the model treats them
separately.

In numerical experiments, artificial social networks were
constructed and a simple greedy algorithm was performed where the
next step in the chain was selected to be the neighbor of the
current node with the smallest distance along any dimension. At
each step in the chain there is a fixed probability, called the
attrition rate, that the node will not pass the message further.
The numerical results showed that for a range of the parameter
$\alpha$ and number of attribute dimensions, the networks are
``searchable'', meaning that a minimum fraction of search paths
find their target.

Kleinberg \cite{kleinberg2000navigation, kleinberg2001dynamics}
posed a related question: in the absence of attrition, when does
the length of the chains scale in the same way as the average
shortest path. Unlike the study of Watts. et al., there is no
attrition - all chains run until completion, but need to scale as
the actual shortest path in the network does. In the case of a
small world network, the average shortest path scales as $\ln(N)$,
where $N$ is the number of nodes. Kleinberg proved that a simple
greedy strategy based on geography could achieve chain lengths
bounded by $(\ln N)^2$ under the following conditions: nodes are
situated on an $m$-dimensional lattice with connections to their
$2*m$ closest neighbors and additional connections are placed
between any two nodes with probability $p \sim r^{-m}$, where $r$
is the distance between them. Since in the real world our
locations are specified primarily by two dimensions, longitude and
latitude, the probability is inversely proportional to the square
of the distance. A person should be four times as likely to know
someone living a block away, than someone two city blocks away.
However, Kleinberg also proved that if the probabilities of
acquaintance do not follow this relationship, nodes would not be
able to use a simple greedy strategy to find the target in
polylogarithmic time.

The models of both Watts et al.\ and Kleinberg show that the
probability of acquaintance needs to be related to the proximity
between individuals' attributes in order for simple search
strategies using only local information to be effective. Below we
describe experiments empirically testing the assumptions and
predictions of the proposed two models.

\subsection{Method}

In order to test the above hypothesis, Adamic and Adar
\cite{adamic03search} applied search algorithms to email networks
derived from the email logs at HP Labs already described in
section \ref{spectroscopy}. A social contact was defined to be
someone with whom an individual had exchanged at least 6 emails
each way over the period of approximately 3 months. The
bidirectionality of the email correspondence guaranteed that a
conversation had gone on between the two individuals and hence
that they are familiar with one another.

 \begin{figure}[tbp]
\begin{center}
\includegraphics[scale=0.40]{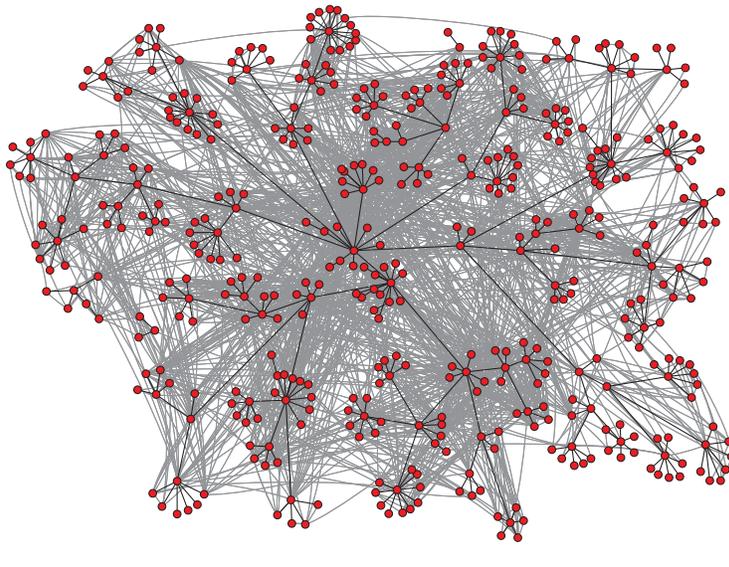}
\end{center}
\caption[Email communications within HP Labs mapped onto the
organizational hierarchy]{Email communications within HP Labs
(gray lines) mapped onto the organizational hierarchy (black
lines). Note that email communication tends to ``cling'' to the
formal organizational chart.} \label{hierarchyemail}
\end{figure}

\begin{figure}[tbp]
\begin{center}
\includegraphics[scale=0.70]{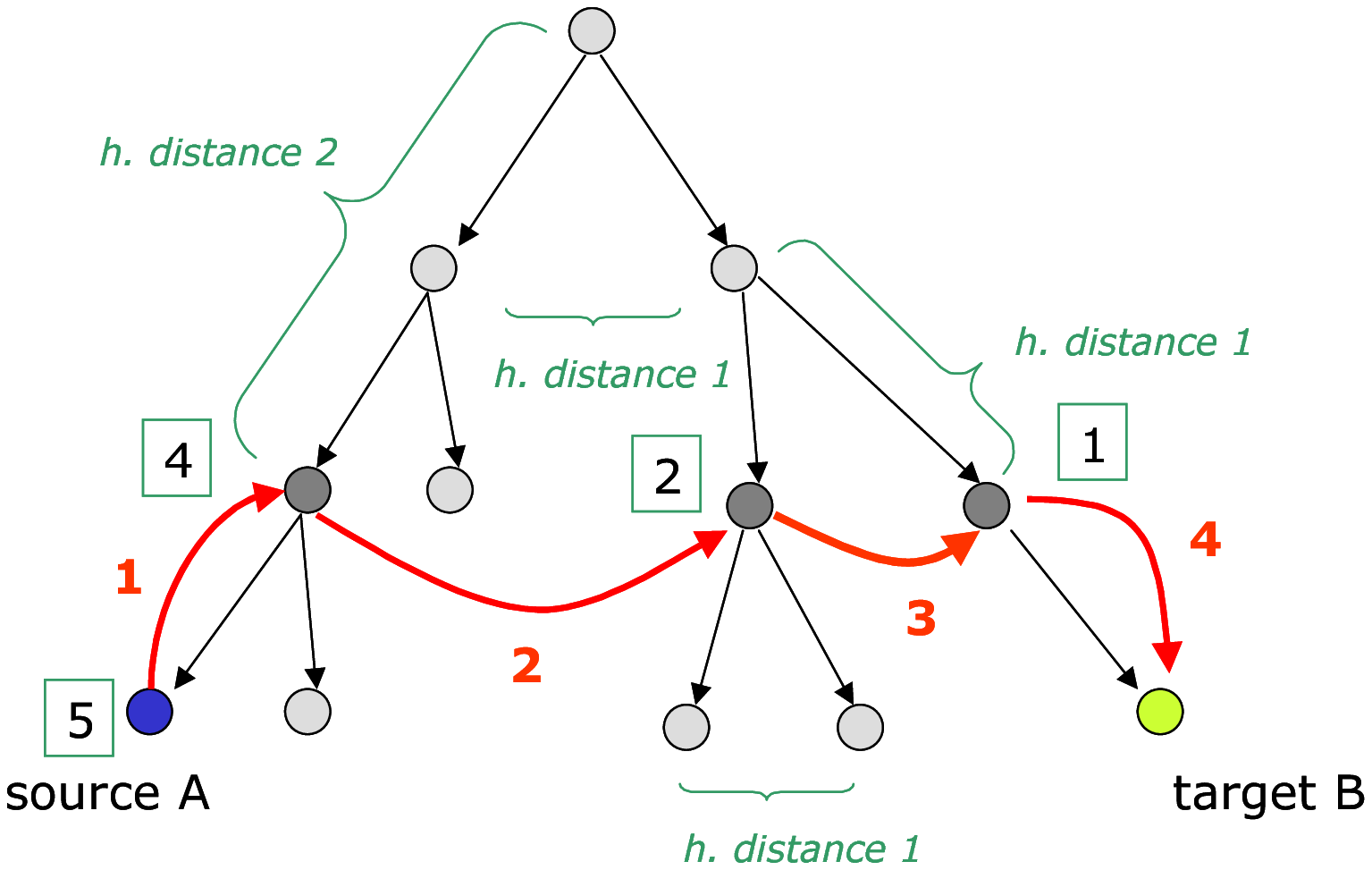}
\end{center}
\caption[Example illustrating a search path using information
about the target's position in the organizational hierarchy
]{Example illustrating a search path using information about the
target's position in the organizational hierarchy to direct a
message. Numbers in the square give the h-distance from the
target.} \label{hierarchyexample}
\end{figure}

Imposing this constraint yielded a network of 436 individuals with
a median number of 10 acquaintances and a mean of 13. The degree
distribution, shown in Figure \ref{swlinkdist}, is highly skewed
with an exponential tail. This is in contrast to the raw power-law
email degree distribution, used in section \ref{infoflow} and
shown in Figure \ref{outdegdist}, pertaining to both internal and
external nodes and possessing no threshold in email volume. A
scale free distribution in the raw network arises because there
are many external nodes emailing just one individual inside the
organization, and there are also some individuals inside the
organization sending out announcements to many people and hence
having a very high degree. However, once we impose a higher cost
for maintaining a social contact (that is, emailing that contact
at least six times and receiving at least as many replies), then
there are few individuals with many contacts.

\subsection{Simulating Milgram's experiment on an email network}

The resulting network, consisting of regular email patterns
between HP Labs employees, had 3.1 edges separating any two
individuals on average, and a median of 3. Simulations were
performed on the network to determine whether members of the
network would be able to use a simple greedy algorithm to locate a
target. In this simple algorithm, each individual can use
knowledge only of their own email contacts, but not their
contacts' contacts, to forward the message.

Three different strategies were tested, at each step passing the
message to the contact who is either
\begin{itemize}
\item best connected

\item closest to the target in the organizational hierarchy

\item sitting in closest physical proximity to the target

\end{itemize}

The first strategy selects the individual who is more likely to
know the target by virtue of the fact that he/she knows so many
people. It has been shown \cite{adamic01plsearch}, that this is an
effective strategy in power-law networks with exponents close to 2
(the case of the unfiltered HP Labs email network), but that it
performs poorly in graphs with a Poisson degree distribution that
has an exponential tail. Since the distribution of contacts in the
filtered HP network was not power-law, the high degree strategy
was not expected to perform well, and this was verified through
simulation. The median number of steps required to find a randomly
chosen target from a random starting point was 17, compared to the
three steps in the average shortest path. Even worse, the average
number of steps is 40. This discrepancy between the mean and the
median is a reflection of the skewness of the distribution: a few
well connected individuals and their contacts are easy to find,
but some individuals who do not have many links and are not
connected to highly connected individuals are difficult to locate
using this strategy.

\begin{figure}[tb]
\begin{center}
\includegraphics[scale=0.5]{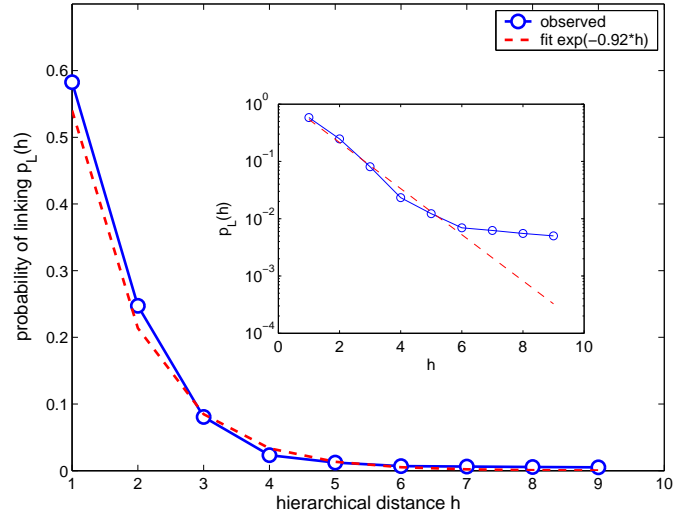}
\end{center}
\caption[Probability of linking as a function of the separation in
the organizational hierarchy]{Probability of linking as a function
of the separation in the organizational hierarchy. The exponential
parameter $\alpha = 0.92$, in the searchable range according to
the model of Watts et al.\cite{watts2002search}
\label{hierarchylinked}}
\end{figure}

The second strategy consisted of passing the message to the
contact closest to the target in the organizational hierarchy. The
strategy relies on the observation, illustrated in Figures
\ref{hierarchyemail} and \ref{hierarchylinked} that individuals
closer together in the organizational hierarchy are more likely to
email with one another. Figure \ref{hierarchyexample} illustrates
such a search, labelling nodes by their hierarchical distance
(h-distance) from the target. The h-distance is computed as
follows: a node has distance one to their manager and to everyone
they share a manager with. Distances are then recursively
assigned, so that each node has h-distance 2 to their first
neighbor's neighbors, and h-distance 3 to their second neighbor's
neighbors, etc. A simple greedy strategy using information about
the organizational hierarchy worked extremely well. The median
number of steps was only 4, close to the median shortest path of
3. With the exception of one individual, whose manager was not
located on site, and who was consequently difficult to locate, the
mean number of steps was 4.7, meaning that not only are people
typically easy to find, but nearly everybody can be found in a
reasonable number of steps.

In the original experiment by Milgram the completed chains were
divided between those that reached the target through his
professional contacts and those that reached him through his
hometown. On average those that relied on geography took 1.5 steps
longer to reach the target, a difference found to be statistically
significant. In the words of Travers and Milgram
\cite{travers69smallworld}, the following seemed to occur:
``Chains which converge on the target principally by using
geographic information reach his hometown or the surrounding areas
readily, but once there often circulate before entering the
target's circle of acquaintances. There is no available
information to narrow the field of potential contacts which an
individual might have within the town.''

\begin{figure}[tb]
\begin{center}
\includegraphics[scale=0.75]{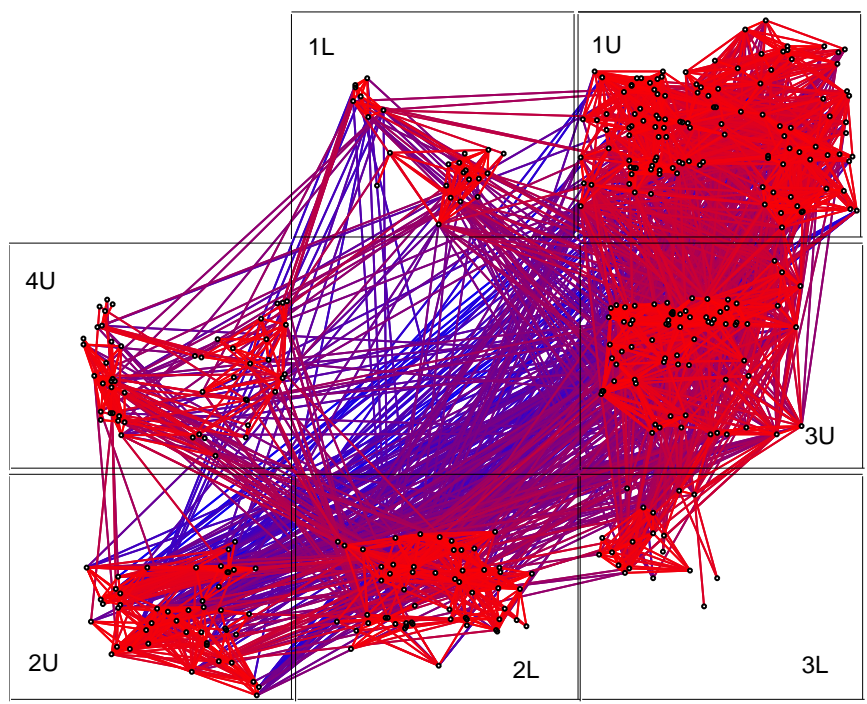}
\end{center}
\caption[Email communications within HP Labs mapped onto
approximate physical location ]{Email communications within HP
Labs mapped onto approximate physical location based on the
nearest post number and building given for each employee. Each box
represents a different floor in a building. The lines are color
coded based on the physical distance between the correspondents:
red for nearby individuals, blue for far away contacts.}
\label{cubiclelayout}
\end{figure}

\begin{figure}[tb]
\begin{center}
\includegraphics[scale=0.5]{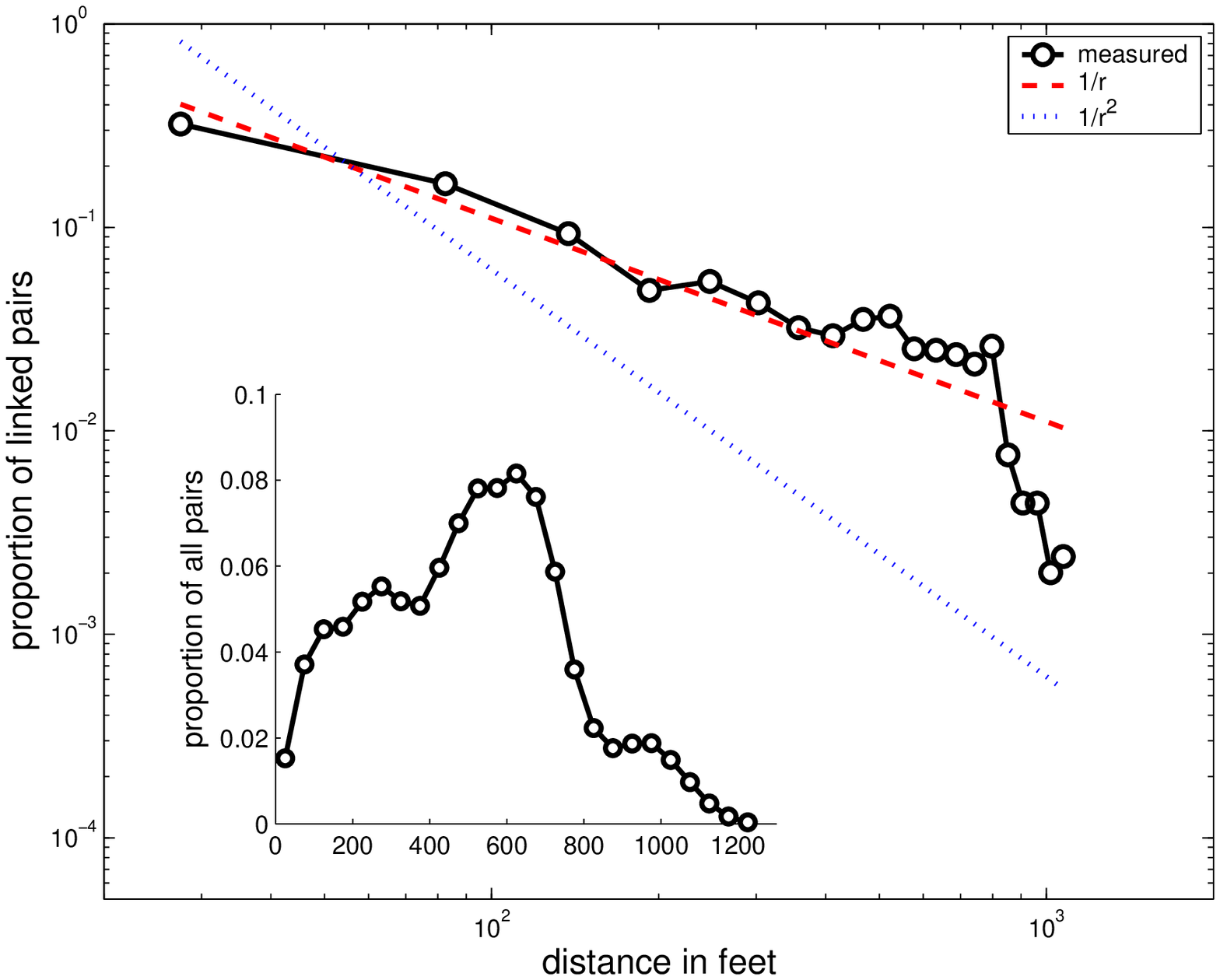}
\end{center}
\caption[Probability of two individuals corresponding by email as
a function of the distance between their cubicles]{Probability of
two individuals corresponding by email as a function of the
distance between their cubicles. The inset shows how many people
in total sit at a given distance from one another.}
\label{cdistandlink}
\end{figure}

Performing the small world experiment on the HP email network
using geography produced a similar result, in that geography could
be used to find most individuals, but was slower, taking a median
number of 7 steps, and a mean of 12. Figure \ref{cubiclelayout}
shows the email correspondence mapped onto the physical layout of
the buildings. Individuals' locations are given by their building,
the floor of the building, and the nearest building post (for
example ``H15'') to their cubicle. The distance between two
cubicles was approximated by the ``street'' distance between their
posts (for example ``A3'' and ``C10'' would be
$(C-A)*25'+(10-3)*25' = 2*25'+7*25' = 225$ feet apart). Adding the
$x$ and $y$ directions separately reflects the interior topology
of the buildings where one navigates perpendicular hallways and
cannot traverse diagonally. If individuals are located on
different floors or in different buildings, the distance between
buildings and the length of the stairway are factored in.

\begin{figure}[tb]
\begin{center}
\includegraphics[scale=0.5]{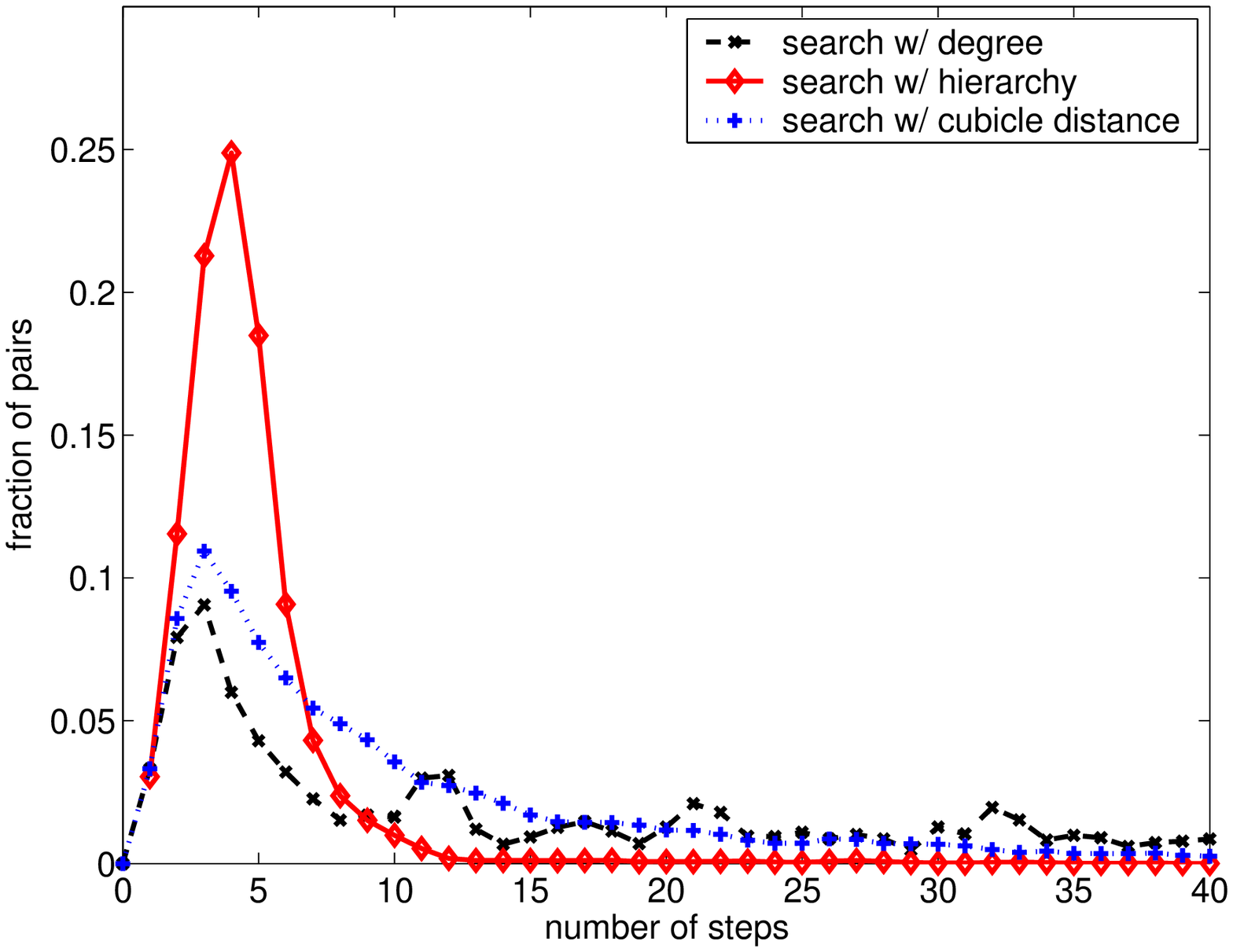}
\end{center}
\caption[Results of simulated search experiments on the HP email
network]{Results of search experiments utilizing either knowledge
of the target's position in the organizational hierarchy or the
physical location of their cubicle.} \label{milgramsearchresults}
\end{figure}

Figure \ref{milgramsearchresults} shows a histogram of chain
lengths resulting from searches using each of the three
strategies. It shows the clear advantage of using the target's
position in organizational hierarchy as opposed to his/her cubicle
location to pass a message through one's email contact. It also
shows that both searches using information about the target
outperform a search relying solely on the connectivity of one's
contacts.

\subsection{Discussion}
The above simulated experiments verify the models proposed in
\cite{watts2002search} and \cite{kleinberg2000navigation} to
explain why individuals are able to successfully complete chains
in the small world experiments using only local information. When
individuals belong to groups based on a hierarchy and are more
likely to interact with individuals within the same small group,
then one can safely adopt a greedy strategy - pass the message
onto the individual most like the target, and they will be more
likely to know the target or someone closer to them.

At the same time it is important to note that the optimum
relationship between the probability of acquaintance and distance
in physical or hierarchical space between two individuals, as
outlined in \cite{kleinberg2000navigation, kleinberg2001dynamics},
are not satisfied. The general tendency of individuals in close
physical proximity to correspond holds: over 87\% percent of the
4000 email links are between individuals on the same floor, and
overall there is a tendency of individuals in close physical
proximity to correspond. Still, individuals maintain
disproportionately many far-flung contacts while not getting to
know some of their close-by neighbors. The relationship between
probability of acquaintance and cubicle distance $r$ between two
individuals, shown in Figure \ref{cdistandlink}, is well-fitted by
a $1/r$ curve. However, Kleinberg has shown that the optimum
relationship in two dimensional space is $1/r^2$ - a stronger
decay in probability of acquaintance than the $1/r$ observed.

\begin{figure}[tb]
\begin{center}
\includegraphics[scale=0.5]{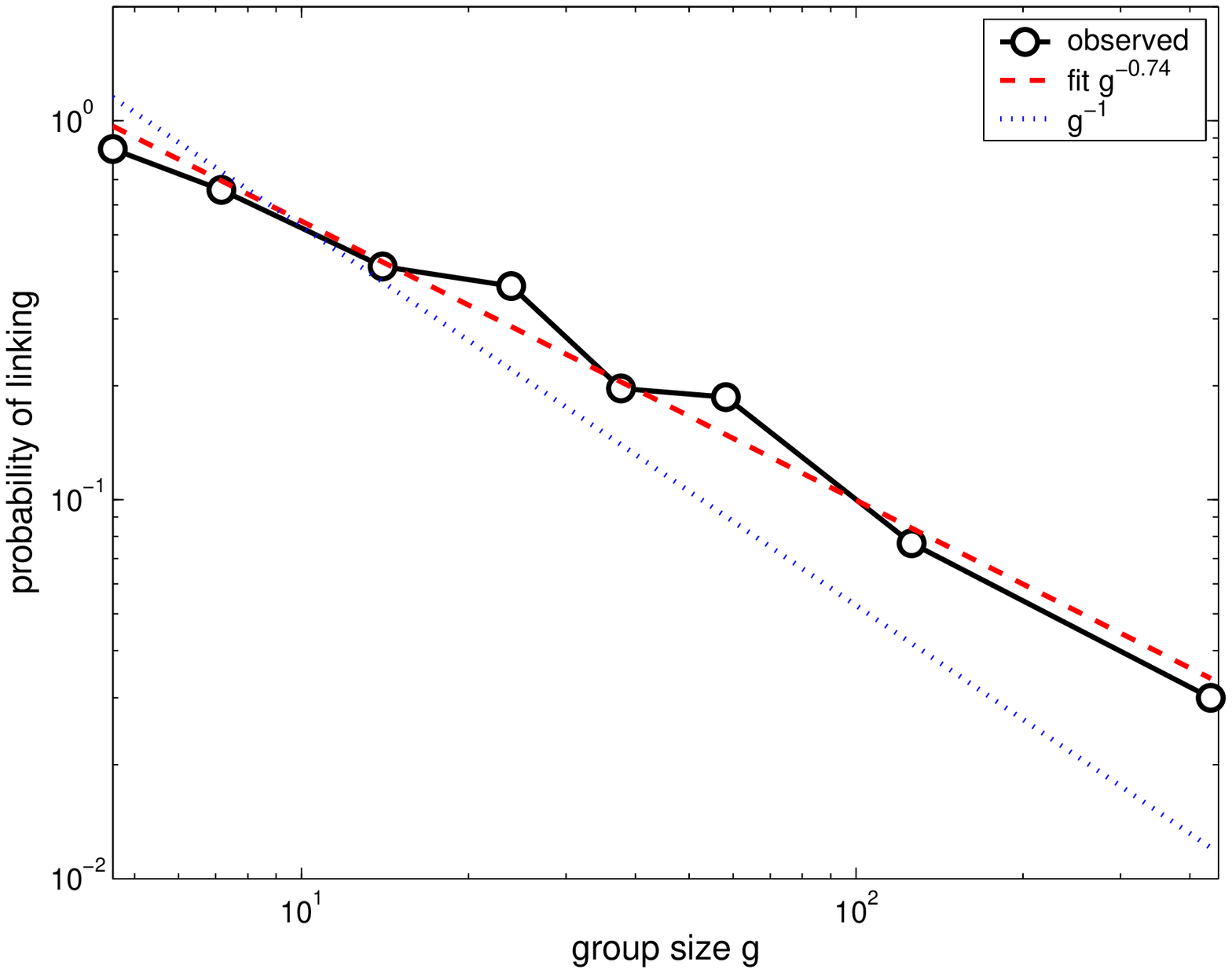}
\end{center}
\caption[Probability of two individuals corresponding by email as
a function of the size of the smallest organizational unit they
both belong to.]{Probability of two individuals corresponding by
email as a function of the size of the smallest organizational
unit they both belong to. The optimum relationship derived in
\cite{kleinberg2001dynamics} is $p \sim g^{-1}$, $g$ being the
group size. The observed relationship is $p \sim g^{-3/4}$. }
\label{groupsizeandlink}
\end{figure}

In the case of HP Labs, the geometry may not be quite two
dimensional, because it is complicated by the particular layout of
the buildings. Hence the optimum relationship may lie between
$1/r$ and $1/r^2$. In any case, the observed $1/r$ probability of
linking shows a tendency consistent with Milgram's observations
about the original small world experiment. At HP Labs, because of
space constraints, re-organizations, and personal preferences,
employees' cubicles may be removed from some of the co-workers
they interact with. This hinders a search strategy relying solely
on geography, because one might get physically quite close to the
target, but still need a number of steps to find an individual who
interacts with them.

The same is true, but to a lesser extent, of the contacts
individuals establish with respect to the organizational
hierarchy. In Section \ref{spectroscopy} email spectroscopy
revealed that while collaborations mostly occurred within the same
organizational unit, they also frequently bridged different parts
of the organization or broke up a single organizational unit into
noninteracting subgroups. The optimum relationship derived in
\cite{kleinberg2001dynamics} for the probability of linking would
be inversely proportional to the size of the smallest
organizational group that both individuals belong to. However, the
observed relationship, shown in Figure \ref{groupsizeandlink} is
slightly off, with $p \sim g^{-3/4}$, $g$ being the group size.

Overall, the results of the email study are consistent with the
model of Watts et al. \cite{watts2002search}. This model does not
require the search to find near optimum paths, but simply
determines when a network is ``searchable'', meaning that fraction
of messages reach the target given a rate of attrition. The
relationship found between separation in the hierarchy and
probability of correspondence, shown in Figure
\ref{hierarchylinked}, is well within the searchable regime
identified in the model.

The study of Adamic and Adar is a first step, validating these
models on a small scale. The email study gives a concrete way of
observing how the small world chains can be constructed. Using a
very simple greedy strategy, individuals across an organization
could reach each other through a short chain of coworkers. It is
quite likely that similar relationships between acquaintance and
proximity (geographical or professional) hold true in general, and
therefore that small world experiments succeed on a grander scale
for the very same reasons.

\section{Conclusion}
In this chapter we reviewed three studies of information flow in
social networks. The first developed a method of analyzing email
communication automatically to expose communities of practice and
their leaders. The second showed that the tendency of individuals
to associate according to common interests influences the way that
information spreads throughout a social group. It spreads quickly
among individuals to whom it is relevant, but unlike a virus, is
unable to infect a population indiscriminately. The third study
showed why small world experiments work - how individuals are able
to take advantage of the structure of social networks to find
short chains of acquaintances. All three studies relied on email
communication to expose the underlying social structure, which
previously may have been difficult and labor-intensive to obtain.
We expect that these findings are also valid with other means of
social communication, such as verbal exchanges, telephony and
instant messenger systems.

\mbox{}

 \textbf{Acknowledgements}

We would like to thank Eytan Adar and TJ Giuli for their comments
and suggestions.

\bibliographystyle{plain}
\bibliography{justinitials}

\begin{thebibliography}{10}

\bibitem{adamic03search}
L.~A. Adamic and E.~Adar.
\newblock How to search a social network.
\newblock submitted for publication,
  \url{http://www.hpl.hp.com/shl/papers/socsearch}, 2002.

\bibitem{adamic03friends}
L.~A. Adamic and E.~Adar.
\newblock Friends and neighbors on the web.
\newblock {\em Social Networks}, 25(3), 2003.

\bibitem{adamic01plsearch}
L.~A. Adamic, R.~M. Lukose, A.~R. Puniyani, and B.~A. Huberman.
\newblock Search in power-law networks.
\newblock {\em Phys. Rev. E}, 64:046135, 2001.

\bibitem{alani02practice}
H.~Alani, K.~O'Hara, and N.~Shadbolt.
\newblock Ontocopi: Methods and tools for identifying communities of practice,
  intelligent information processing conference.
\newblock In {\em IFIP World Computer Congress (WCC)}, 2002.

\bibitem{allen84flow}
T.~Allen.
\newblock {\em Managing the Flow of Technology}.
\newblock MIT Press, Cambrige, MA, 1984.

\bibitem{blau63organizations}
P.M. Blau and W.R. Scott.
\newblock {\em Formal organizations. A comparative approach}.
\newblock Lndn. Routledge \& Kegan Paul, 1963.

\bibitem{brandes01betweenness}
U.~Brandes.
\newblock A faster algorithm for betweenness centrality.
\newblock {\em Journal of Mathematical Sociology}, 25(2):163--177, 2001.

\bibitem{burt80network}
R.~S. Burt.
\newblock Models of network structure.
\newblock {\em Annual Review of Sociology}, 6:79--141, 1980.

\bibitem{crane72colleges}
D.~Crane.
\newblock {\em Invisible Colleges: Diffusion of Knowledge in Scientific
  Communities}.
\newblock University of Chicago Press, Chicago, 1972.

\bibitem{crozier64bureaucratic}
M.~Crozier.
\newblock {\em The Bureaucratic Phenomenon}.
\newblock University of Chicago Press, Chicago, 1964.

\bibitem{zoltan02halting}
Z.~Dezso and A.-L. Barabasi.
\newblock Halting viruses in scale-free networks.
\newblock {\em Phys. Rev. E}, 65:055103, 2002.

\bibitem{dodds03networks}
P.~S. Dodds, R.~M., and D.~J. Watts.
\newblock An experimental study of search in global social networks.
\newblock {\em Science}, 301:827--829, 2003.

\bibitem{ducheneaut02email}
N.~Ducheneaut and V.~Belloti.
\newblock A study of email work processes in three organizations.
\newblock to appear in the Journal of CSCW, 2002.

\bibitem{ebel02email}
H.~Ebel, L.-I. Mielsch, and S.~Bornholdt.
\newblock Scale-free topology of e-mail networks.
\newblock {\em Phys. Rev. E}, 66:035103, 2002.

\bibitem{eckmann03email}
J.-P. Eckmann, E.~Moses, and D.~Sergi.
\newblock Dialog in e-mail traffic.
\newblock \url{http://xyz.lanl.gov/abs/cond-mat/0304433"}, 2003.

\bibitem{eguiluz02epidemicclust}
V.~M. Eguiluz and K.~Klemm.
\newblock Epidemic threshold in structured scale-free networks.
\newblock {\em Phys. Rev. Lett.}, 89:108701, 2002.

\bibitem{feld81social}
S.L. Feld.
\newblock The focused organization of social ties.
\newblock {\em American Journal of Sociology}, 86:1015--1035, 1981.

\bibitem{freeman77centrality}
L.~Freeman.
\newblock A set of measures of centrality based on betweenness.
\newblock {\em Sociometry}, 40:35--41, 1977.

\bibitem{fruchterman91graph}
T.~M.~J. Fruchterman and E.~M. Reingold.
\newblock Graph drawing by force-directed placement.
\newblock {\em Software - Practice and Experience}, 21(11):1129--1164, 1991.

\bibitem{garton97online}
L.~Garton, C.~Haythornwaite, and B.~Wellman.
\newblock Studying on-line social networks.
\newblock {\em Journal of Computer Mediated Communication}, 3(1), 1997.

\bibitem{girvan02community}
M.~Girvan and M.E.J. Newman.
\newblock Community structure in social and biological networks.
\newblock {\em Proc. Natl. Acad. Sci. USA}, 99:8271--8276, 2002.

\bibitem{guimera02email}
R.~Guimer\`{a}, L.~Danon, A.~D\'{i}az-Guilera, F.~Giralt, and A.~Arenas.
\newblock Self-similar community structure in organizations.
\newblock \url{http://arxiv.org/PS_cache/cond-mat/pdf/0211/0211498.pdf}, 2002.

\bibitem{hinds95communication}
P.~Hinds and S.~Kiesler.
\newblock Communication across boundaries: Work, structure, and use of
  communication technologies in a large organization.
\newblock {\em Organization Science}, 6(4):373--393, 1995.

\bibitem{huberman95communities}
B.A. Huberman and T.~Hogg.
\newblock Communities of practice: Performance and evolution.
\newblock {\em Computational and Mathematical Organization Theory}, 1:73--92,
  1995.

\bibitem{killworth78reverse}
P.~Killworth and H.~Bernard.
\newblock Reverse small world experiment.
\newblock {\em Social Networks}, 1:159--192, 1978.

\bibitem{kleinberg2000navigation}
J.~Kleinberg.
\newblock Navigation in a small world.
\newblock {\em Nature}, 406, 2000.

\bibitem{kleinberg2001dynamics}
J.~Kleinberg.
\newblock Small-world phenomena and the dynamics of information.
\newblock {\em Advances in Neural Information Processing Systems (NIPS)}, 14,
  2001.

\bibitem{krebs02terrorist}
V.~E. Krebs.
\newblock Uncloaking terrorist networks.
\newblock {\em First Monday}, 7(4), April 2002.

\bibitem{lave91learning}
J.~Lave and E.~Wenger.
\newblock {\em Situated Learning: Legitimate Peripheral Participation}.
\newblock Cambridge University Press, 1991.

\bibitem{lazarsfeld54friendship}
P.~Lazarsfeld and R.K.Merton.
\newblock In M.~Berger, T.~Abel, and C.H. Page, editors, {\em Freedom and
  Control in Modern Society}, chapter Friendship as a social Process: A
  Substantive and Methodological Analysis. Van Nostrand, New York, 1954.

\bibitem{lundberg75smallworld}
C.~C. Lundberg.
\newblock Patterns of acquaintanceship in society and complex organization: A
  comparative study of the small world problem.
\newblock {\em Pacific Sociological Review}, 18:206--222, 1975.

\bibitem{milgram67}
S.~Milgram.
\newblock The small-world problem.
\newblock {\em Psychology Today}, 1:62--67, 1967.

\bibitem{newman02assortative}
M.~E.~J. Newman.
\newblock Assortative mixing in networks.
\newblock {\em Phys. Rev. Lett}, 89:208701, 2002.

\bibitem{newman02emailnetworks}
M.~E.~J. Newman, S.~F., and J.~Balthrop.
\newblock Email networks and the spread of computer viruses.
\newblock {\em Phys. Rev. E}, 66:035101, 2002.

\bibitem{newman01graphs}
M.~E.~J. Newman, S.~H. Strogatz, and D.~J. Watts.
\newblock Random graphs with arbitrary degree distribution and their
  applications.
\newblock {\em Phys. Rev. E}, 64:026118, 2001.

\bibitem{newman01scientist}
M.E.J. Newman.
\newblock Who is the best connected scientist? a study of scientific
  coauthorship networks.
\newblock {\em Phys. Rev. E}, 64:016131, 2001.

\bibitem{newman02epidemic}
M.E.J Newman.
\newblock The spread of epidemic disease on networks.
\newblock {\em Phys. Rev. E}, 66:016128, 2002.

\bibitem{ouchi80markets}
W.~G. Ouchi.
\newblock Markets, bureaucracies, and clans.
\newblock {\em Administrative Science Quarterly}, 25:129--141, 1980.

\bibitem{pastor-satorras01epidemic}
R.~Pastor-Satorras and A.~Vespignani.
\newblock Epidemic spreading in scale-free networks.
\newblock {\em Phys. Rev. Lett.}, 86(14):3200--3203, 2001.

\bibitem{pool78contacts}
I.~Pool and M.~Kochen.
\newblock Contacts and influence.
\newblock {\em Social Networks}, 1:5--51, 1978.

\bibitem{schwartz93discovering}
M.~F. Schwartz and D.~C.~M. Wood.
\newblock Discovering shared interests among people using graph analysis.
\newblock {\em Communications of the ACM}, 36(8):78--89, 1993.

\bibitem{scott92organizations}
W.~R. Scott.
\newblock {\em Organizations: Rational, Natural, and Open Systems}.
\newblock Prentice-Hall, Englewood Cliffs, NJ, 1992.

\bibitem{sparrow91criminal}
M.~K. Sparrow.
\newblock The application of network analysis to criminal intelligence: An
  assessment of the prospects.
\newblock {\em Social Networks}, 13:251--274, 1991.

\bibitem{touhey74similarity}
J.C. Touhey.
\newblock Situated identities, attitude similarity, and interpersonal
  attraction.
\newblock {\em Sociometry}, 37:363--374, 1974.

\bibitem{travers69smallworld}
J.~Traver and S.~Milgram.
\newblock An experimental study of the small world problem.
\newblock {\em Sociometry}, 32:425--443, 1969.

\bibitem{tyler03email}
J.~R. Tyler, D.~M. Wilkinson, and B.~A. Huberman.
\newblock Email as spectroscopy: Automated discovery of community structure
  within organizations.
\newblock In {\em Proceedings of the International Conference on Communities
  and Technologies}. Kluwer Academic Publishers, Netherlands, 2003.

\bibitem{vazquezPRE2003}
A.~Vazquez, M.~Boguna, Y.~Moreno, R.~Pastor-Satorras, and A.~Vespignani.
\newblock Topology and correlations in structured scale-free networks.
\newblock {\em Physical Review E}, 67:046111, 2003.

\bibitem{wasserman94socialnetwork}
S.~Wasserman and K.~Faust.
\newblock {\em Social network analysis}.
\newblock Cambridge University Press, Cambridge, 1994.

\bibitem{watts2002search}
D.~J. Watts, P.~S. Dodds, and M.~E.~J. Newman.
\newblock Identity and search in social networks.
\newblock {\em Science}, 296:1302--1305, 2002.

\bibitem{watts98smallworld}
D.~J. Watts and S.~H. Strogatz.
\newblock Collective dynamics of small-world networks.
\newblock {\em Nature}, 393:440--442, 1998.

\bibitem{wellman02computersocial}
B.~Wellman.
\newblock Computer networks as social networks.
\newblock {\em Science}, 293:2031--34, 2002.

\bibitem{whittaker96email}
S.~Whittaker and C.~Sidner.
\newblock Email overload: exploring personal information management of email.
\newblock In {\em Proceedings of CHI'96 Conference on Computer Human
  Interaction}, pages 276--283. Logos Verlag, New York, 21996.

\bibitem{wilkinson02genes}
D.~Wilkinson and B.~A. Huberman.
\newblock A method for finding communities of related genes.
\newblock submitted for publication,
  \url{http://www.hpl.hp.com/shl/papers/communities/index.html}, 2002.

\bibitem{wu03flow}
F.~Wu, B.~A. Huberman, L.~A. Adamic, and J.R. Tyler.
\newblock Information flow in social groups.
\newblock \url{http://arxiv.org/abs/cond-mat/0305305"}, 2003.

\end{thebibliography}
\end{document}